\documentclass{elsart}
\usepackage{graphics,graphicx,amssymb}

\newbox\grsign \setbox\grsign=\hbox{$>$} \newdimen\grdimen \grdimen=\ht\grsign
\newbox\simlessbox \newbox\simgreatbox
\setbox\simgreatbox=\hbox{\raise.5ex\hbox{$>$}\llap
     {\lower.5ex\hbox{$\sim$}}}\ht1=\grdimen\dp1=0pt
\setbox\simlessbox=\hbox{\raise.5ex\hbox{$<$}\llap
     {\lower.5ex\hbox{$\sim$}}}\ht2=\grdimen\dp2=0pt

\begin{document}

\begin{frontmatter}

\title{Inequality reversal: effects of the savings propensity and correlated returns}

\large{Anindya S. Chakrabarti$~^{1*}$, Bikas K. Chakrabarti$~^{2\dag}$}

\textit{\small{$~^1$ Indian Statistical Institute, \\203 B. T. Road, Kolkata 700018, India}}

\textit{\small{$~^2$ ERU, Indian Statistical Institue, \\203 B. T. Road, Kolkata 700018, India\\
and\\
CAMCS,
Saha Institute of Nuclear Physics,\\
1/AF Bidhannagar, Kolkata 700064, India}}

\thanks{Email addresses: $~^*$aschakrabarti@gmail.com (corresponding author), $~^\dag$bikask.chakrabarti@saha.ac.in}

\begin{abstract}
\noindent 
\noindent In the last decade, a large body of literature has been developed to explain the universal
features of inequality in terms of income and wealth. By now, it is established that the distributions of
income and wealth in various economies 
show a number of statistical regularities. There are several models to explain such
static features of inequality
in an unifying framework and the kinetic exchange models, in particular, provide one such framework.  
Here we focus on the dynamic features of inequality.
In the process of development and growth, inequality in an economy in terms of income and wealth
follows a particular pattern of rising in the initial stage followed by an eventual fall. This inverted
U-shaped curve is known
as the Kuznets Curve. We
examine the possibilities of such behavior of an economy in the context of a generalized kinetic exchange
model. It is shown that under some specific conditions, our model economy indeed shows
inequality reversal.

\end{abstract}
\end{frontmatter}

\section{Introduction} 
\noindent 

The distributions of income
and wealth have long been found to possess some robust and stable
features independent of the specific economic, social and political conditions of the economies.
Traditionally, the economists have preferred to model the left tail and the mode of the distributions of 
the workers' incomes
with a log-normal 
distribution and the heavier right tail with a Pareto distribution. For incomes from assets, other functional forms 
have been used as models, e.g., the family of functions introduced by Camilo Dagum \cite{dagum}.
For a detailed survey of the distributions used to fit the income and wealth data see Ref. \cite{kleiber}.
However, there have been several studies recently that argue that
the left tail and the mode of
the distribution fit well with the gamma distribution and the right tail of the distribution follows a power
law \cite{acbkc;07}. It has been argued that this feature might be considered to be a natural law
for economics \cite{yako-rosser;09}.
There is yet another well known observation, represented by the so-called Kuznets Curve, that
over time a growing economy shows a rise in inequality followed by the eventual 
fall \cite{nielsen;94}.

The standard explanation of the Kuznets Curve \cite{kuznets;55,kuznets;65} goes like the following. 
Suppose, there is an economically underdeveloped area {\it A}, one
in which people live from pre-industrial farming, perhaps not even involved in much of a monetized market economy.
They generate little wealth and less money.
Then a small part of {\it A}, say {\it B}, undergoes initial industrialization generating more wealth and 
creating a small monetized market economy. Hence, {\it A}'s inequality as a whole increases because those who
are in {\it B}, enjoy a larger amount of wealth and the rest still remain the same. However, as more and more 
industrialization takes place, a larger fraction of {\it A}'s people come into {\it B} and hence, inequality falls.
It is clear that this explanation rests on the effects of migration and growth of an economy during the process of 
industrialization.

However, in this paper, we examine the possibility of such inequality reversals in a conservative
kinetic exchange model 
of market economy where no growth or migration takes place.
Angle {\it et al} \cite{angle;10} made one of the first attempts to explain the Kuznets Curve by a kinetic
exchange model. They assumed two populations
characterized by two different gamma distributions and by incorporating the effects of `education' and `purchaing
power of money', they had 
shown the possibility of Kuznets Curve in such an economy.
Here, we consider only a single population of constant size.  
A very simple binary trading process based on a micro-economic framework, is modeled. Later, we show that
the same process essentially captures the idea of the market returns being correlated.
Next, it is shown that this model is very general as it reproduces
a number of basic ideal gas like market models in certain limits of its parameters and the model becomes useful
as it allows us to examine the effects of the variations in the savings propensity and the correlation in market returns
explicitly.

We then study the dynamic behavior of the steady state distributions of asset with changes in the savings propensity
and the correlation parameter.
We show that in this model economy, the incorporation of the above two phenomena can produce the
`inequality reversal' in terms of asset-holding.

This paper is organized as follows.
In section \ref{sec:exchange}, we propose a particular kind of binary trading process in a competitive market. In the
next section, we propose a model which enables us to examine the effects of correlation in market returns when
the agents trade in several markets simultaneously. It is shown that same binary trading process can be used to capture 
the effects of correlation in market returns
as well. In section \ref{sec:model}, we study the derived kinetic
exchange model and in the next section, we derive the Kuznets Curve. 
Then follows a summary.

\section{The Asset Exchange Equations}
\label{sec:exchange}

\medskip

\noindent Borrowing the framework studied in Ref. \cite{chakrabartis;09}, we consider an $N$-agent exchange economy.
Each of the agents produce a single
non-storable good and each of the goods is different from another. Money is treated as another commodity.
However, the model is conservative in that money is neither created nor destroyed. No agent dies and none
is born.
Money
facilitates transactions in this model economy. Time is discrete.
The agents care for their future consumptions and hence they care
about their savings in the current period as well. Each of these agents
are endowed with an initial amount of money which we assume to be unity
for every agent for simplicity.
At each time step, two agents are chosen randomly and they carry out transactions according
to their utility maximization principle in a general equilibrium set up i.e., prices are so determined
that the demands of the goods are exactly matched by the supplies, clearing the markets.
We assume that
the parameters of the utility function can vary over time \cite{chakrabartis;09, jet;02} reflecting the fact that the
agents' preferences alter with time. Ref. \cite{jet;02} presented a similar framework in a different context.

Suppose agent 1 produces $Q_1$ amount of commodity 1 only and agent 2 produces
$Q_2$ amount of
commodity 2 only and the amounts of money in their possession at
time $t$ are $m_1(t)$ and $m_2(t)$ respectively.
In this scenario, both of them would be willing to trade and buy the
other good by selling a fraction of their own productions
as well as with the money that they hold. Hence, at each time step
there would be a net transfer
of money from one agent to the other due to such trade.
For notational convenience, we denote $m_i(t+1)$ as $m_i$
and $m_i(t)$ as $M_i$ (for $i=1, 2$).
We define the utility functions as follows.
For agent $1$, {$U_1(x_1,x_2,m_1)= x_1^{\alpha_1}x_2^{\alpha_2}m_1^{\lambda}$}
and for agent $2$,
$U_2(y_1,y_2,m_2)=y_1^{\beta_1}y_2^{\beta_2}m_2^{\lambda}$.
The arguments in both of the utility functions are consumption of the first (i.e.,
$x_1$ and $y_1$) and second good (i.e., $x_2$ and $y_2$)
and amount of money in their possession respectively.
For simplicity, we normalize the sum of the powers to $1$ i.e.,
$\alpha_1+\alpha_2+\lambda=1$ and $\beta_1+\beta_2+\lambda=1$.
Let the market clearing prices be denoted by $p_1$ and $p_2$.
Note that money acts as the {\it numeraire} good. Hence, its price is unity.
Now, we can define the budget constraints as follows. For agent $1$
the budget constraint is $p_1x_1+p_2x_2+m_1\leq M_1+p_1Q_1$ and
similarly, for agent $2$ the constraint is
$p_1y_1+p_2y_2+m_2 \leq M_2+p_2Q_2$.

\noindent {\bf Proposition}: {\it In such a competitive market with binary trading between agents 
characterized by the above-mentioned
Cobb-Dauglas type utility function, the asset exchange equations will contain two correlated random variables.}

\medskip

\noindent {\it Proof}: Formally, agent 1's problem is to maximize $U_1(x_1,x_2,m_1)=
x_1^{\alpha_1}x_2^{\alpha_2}m_1^{\lambda}$ subject to the budget constraint $p_1x_1+p_2x_2+m_1= M_1+p_1Q_1$
and for agent $2$, the problem is to maximize
$U_2(y_1,y_2,m_2)=y_1^{\beta_1}y_2^{\beta_2}m_2^{\lambda}$
subject to the constraint
$p_1y_1+p_2y_2+m_2= M_2+p_2Q_2$.

Let us solve the problem for the first agent by using Lagrange multiplier technique.
\begin{equation}
{\mathcal L}=x_1^{\alpha_1}x_2^{\alpha_2}m_1^{\lambda} -\mu(p_1x_1+p_2x_2+m_1- M_1+p_1Q_1) \nonumber
\end{equation}
\noindent Equating the first derivatives (with respect to $x_1$, $x_2$, $m_1$ and $\mu$) with zero,
one can derive the demand functions of the first agent as the following.
$$x^*_1=\alpha_1\frac{(M_1+p_1Q_1)}{p_1},~ x^*_2=\alpha_2\frac{(M_1+p_1Q_1)}{p_2},$$
$$m^*_1=\lambda(M_1+p_1Q_1).$$
\noindent Similarly for agent 2, the demand functions are
$$y^*_1=\beta_1\frac{(M_2+p_2Q_2)}{p_1},~y^*_2=\beta_2\frac{(M_2+p_2Q_2)}{p_2}, $$
$$m^*_2=\lambda(M_2+p_2Q_2).$$

\noindent The market clearing conditions are
$x^*_1+y^*_1=Q_1$ and $x^*_2+y^*_2=Q_2$ (i.e., demand is exactly matched by supply in both the markets).
By substituting the values of $x^*_1$, $x^*_2$, $y^*_1$ and $y^*_2$ and
by solving these two equations we get market clearing prices
($\hat p_1$, $\hat p_2$) where
$$\hat p_1=\frac{\left(\lambda\alpha_1+\beta_1(1-\lambda)\right)M_1+\beta_1 M_2}{\lambda Q_1(1-\alpha_1+\beta_1)}$$
\noindent and
$$\hat p_2=\frac{\alpha_2 M_1+\left((1-\lambda)\alpha_2+\lambda\beta_2\right)M_2}{\lambda Q_1(1-\alpha_1+\beta_1)}.$$
\noindent By substituting ($\hat p_1$, $\hat p_2$) in the money demand equations, we get

\begin{eqnarray}
m_1^* &=& \lambda M_1 +\frac{\lambda\alpha_1+(1-\lambda)\beta_1}{1-\alpha_1+\beta_1} M_1
+\frac{\beta_1}{1-\alpha_1+\beta_1}M_2 \nonumber\\
m_2^* &=& \lambda M_2(t)+\frac{\alpha_2}{1-\alpha_1+\beta_1}M_1 
+\frac{\lambda\beta_2+(1-\lambda)\alpha_2}{1-\alpha_1+\beta_1} M_2.
\end{eqnarray}
\noindent Now, we denote
$m_i$ as $m_i(t+1)$
and $M_i$ as $m_i(t)$ (for $i=1, 2$).

\noindent Note that by assuming $\alpha_i=\beta_i$ (for $i$=$1$, $2$) one can derive the CC model \cite{anirbanc;00}
(see Ref. \cite{chakrabartis;09} for the derivation of the CC model). The above set of equations can be rewritten as
\begin{eqnarray}
m_1(t+1) &=& \lambda m_1(t) +\theta_{11} m_1(t) +\theta_{12} m_2(t) \nonumber\\
m_2(t+1) &=& \lambda m_2(t) +\theta_{21} m_1(t) +\theta_{22} m_2(t)
\label{theta}
\end{eqnarray}
\noindent by appropriately defining $\theta_{ij}$s (for $i,j$=$1,2$).
One can very easily verify that the total amount of money remains conserved at each trading i.e.,
$$m_1(t+1)+m_2(t+1)=m_1(t)+m_2(t).$$
\noindent The presence of a positive savings propensity is evident in Eqn. \ref{theta}.
Assuming that $\alpha_i$ and $\beta_i$ (for $i$=1, 2) are random variables (because of the
time dependence of preference
ordering), we see that
the $\theta_{ij}$s are correlated (for $i,j$=1, 2). Hence, the money transfer equations consists of two
correlated random terms.        \\ $~~~~~~~~~~~~~~~~~~~~~~~~~~~~~~~~~~~~~~~~~~~~~~~~~~~~~~~~~~~~~~~~~~~~~~~~~~~$ 
$~~~~~~~~~~~~~~~~~~~~~~~$                  $\square$     \\

For simplicity we assume the $\theta_{ij}$s to be correlated in the following form. 

\begin{eqnarray}
m_i(t+1) &=& \lambda m_i(t)+\omega_1(1-\lambda) m_i(t)+\left(\alpha\omega_1+(1-\alpha)\omega_2\right) \nonumber \\
&& (1-\lambda) m_j(t) \nonumber\\
m_j(t+1) &=& \lambda m_j(t)+(1-\omega_1)(1-\lambda)m_i(t)+\left(1-\alpha\omega_1-(1-\alpha)\omega_2\right) \nonumber \\
&&(1-\lambda) m_j(t)
\label{model}
\end{eqnarray}

where $\omega_1$, $\omega_2$ $\sim$ uniform[0, 1] and independent.
Note that in the above formulation
$\omega_1$ and $(\alpha\omega_1+(1-\alpha)\omega_2)$ are the two correlated random terms.
Later, in section \ref{sec:ineq}, we shall
see that the distributional assumptions of $\omega_1$ and $\omega_2$ will help us to calculate the
moments of the resultant steady state distributions of money, very easily.
The savings propensity
and the degree of correlation between the stochastic terms are denoted
by $\lambda$ and $\alpha$ respectively and both can vary between $0$ and $1$.
Technically, $\alpha$ is not the {\it correlation coefficient}. It is a term that
helps us to tune the correlation betweeen the two random terms.

While deriving the above model, we have assumed that the agents are producing only one commodity at every time-point.
However, we can present the same model by incorporating the idea of risk-aversion 
explicitly where each agent produces a vector of commodities. Below we discuss the notion of correlation in the returns
from trading of several commodities simultaneously.

\section{Correlated Markets}
\label{sec:corrmkt}
\noindent 
Let us begin with a simple calculation.
Suppose, an agent invests a certain amount of money in $K$ number of assets where the returns are stochastic. More
precisely, let us assume that the returns ($\epsilon_k$) are i.i.d. variables with
finite mean ($\mu$) and variance ($\sigma^2$). The problem of
the agent is to decide what fractions ($f_k$) of his money holding he would invest in each asset $k$ for $k=1,2,...,K$.
Assuming risk aversion, the problem is to minimize the variance of his portfolio ($\sum_{k}\epsilon_kf_k $) or formally, the problem is
to minimize $$\sigma^2(\epsilon)\left(f_1^2+f_2^2+...+f_K^2\right) \nonumber $$ 
subject to the condition that $$f_1+f_2+...+f_K=1.  \nonumber $$ 
\noindent Clearly,
the solution would be $f_k^*=1/K$ for all $k$. 
Now, consider the following set of stochastic difference equations representing how the money-holding changes among the
agents over time.

\begin{eqnarray}
m_i(t+1) &=& \epsilon[m_i(t)+m_j(t)] \nonumber\\
\label{randshare}
m_j(t+1) &=& (1-\epsilon)[m_i(t)+m_j(t)]   \nonumber\\
\end{eqnarray}

This framework had been borrowed from the statistical mechanics of the scattering process and numerous
variations of Eqn. \ref{randshare} have been studied in the literature, in the context of income 
and wealth distributions (see e.g., Ref. \cite{yako-rosser;09, chakrabartis;10}). However,
Ref. \cite{chakrabartis;09} presents a microeconomic model in which the same set of equations is obtained
by the market-clearing trading process between the agents. There it had been assumed that each of the agents 
produced a
single non-storable commodity and money acted as an asset that helps to make transactions (the same assumptions
have been made in the last section while deriving the binary trading equations). However, we can generalize
the situation
assuming that each of the agents produce a vector of commodities and engage in trading with each other, then
it is perfectly possible for a risk averse agent to diversify his money holding
at time $t$ following the above calculation, instead of putting all his money in trading of a single commodity. 
The market has the following structure. Each agent produces $K$ ($K\ge1$) number of commodities and each
of these commodities is different from the other. Hence, the agents would be willing to trade with each other.
Ref. \cite{chakrabartis;09} deals with the case where $K=1$ i.e., each agent produces a single commodity and 
it shows that Eqn. \ref{randshare} captures the basic process of money exchange in such an economy. 
Here, we consider the case where $K\ge1$.
Clearly, the risk averse agents would diversify their money holding in order to minimize the risk from trading.
The mode of trading is such that at each instant, two randomly chosen agents engage in trading each producing 
$K$ number of different goods so that, in total, $2K$ number of goods are traded at each instant.
For trading the $k$-th pair of goods ($k=1,2,...,K$), the $i$-th and the $j$-th
agent uses $m_i/K$ and $m_j/K$ amounts of money respectively because we
have already shown that the variance (risk) minimizing choice is to diversify equally among all assets.
So the money transfer equations become the generalisation of Eqn. \ref{randshare}, viz.,
\begin{eqnarray}
m_i(t+1)=\frac{\sum_{k}\epsilon_k}{K}[m_i(t)+m_j(t)] \nonumber \\
\label{diversify}
m_j(t+1)=\left(1-\frac{\sum_{k}\epsilon_k}{K}\right)[m_i(t)+m_j(t)]  \nonumber \\
\end{eqnarray}
\noindent for all possible integer values of $K$.

\begin{figure}
\begin{center}
\noindent \includegraphics[clip,width= 6cm, angle = 270]
{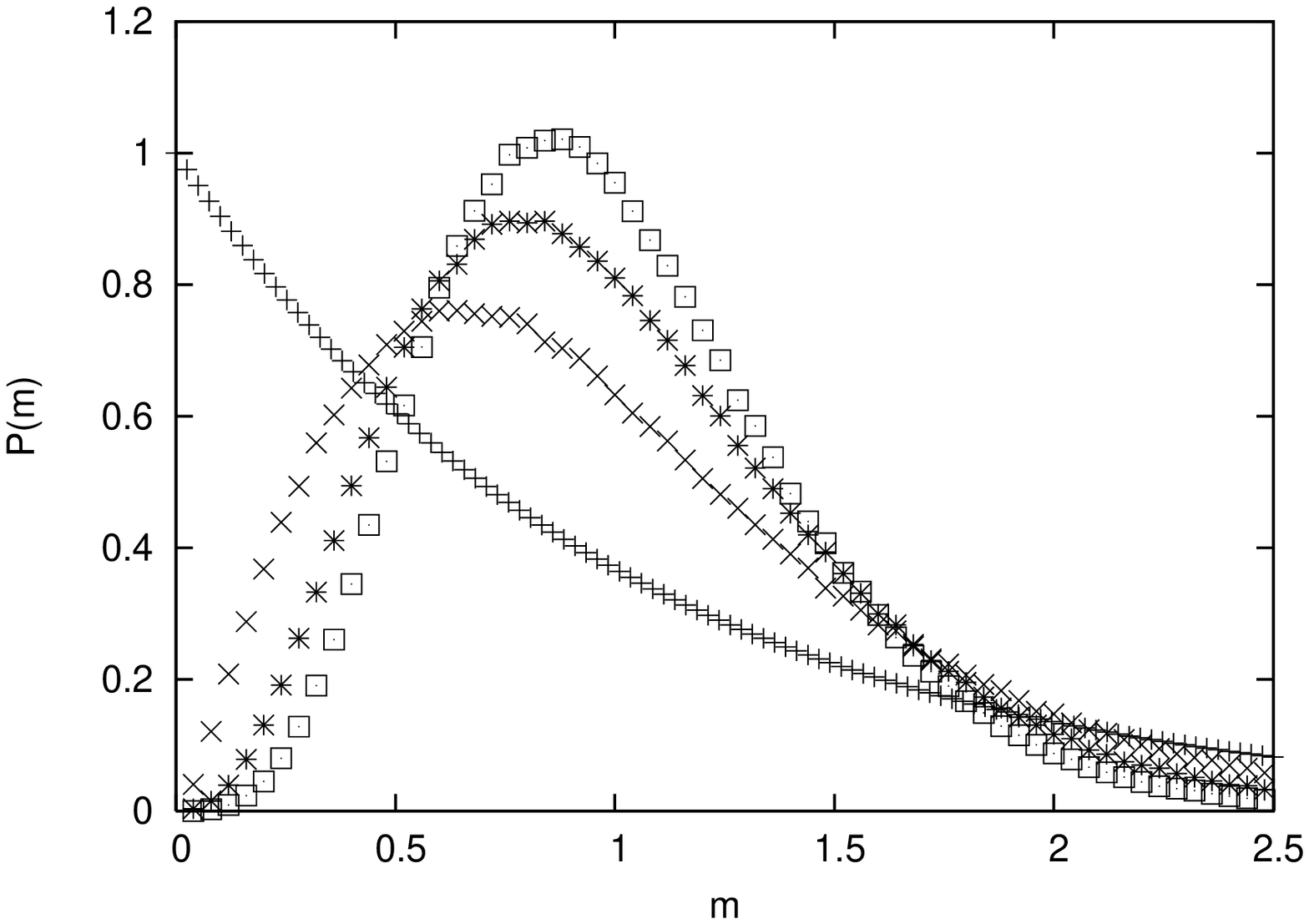}
\caption\protect{Money distribution among risk-averse agents. Four cases are shown
above, viz., $K=1$ (+), $K=2$ ($\times$),
$K=3$ ($\ *$), $K=4$ ($\square$).
All simulations are done
for $O(10^6)$ time steps with 100 agents and averaged over $O(10^3)$ time steps.

}
\label{div}
\end{center}
\end{figure}

{\bf Corollary}: If $K$ is $1$, then we get back Eqn. \ref{randshare}
which implies that the steady state distribution of money would be exponential.
In the other extreme for $lim~~K\rightarrow\infty$,
by applying Lindeberg-Levy central limit theorem, we have
$$\left(\frac{\Sigma_k\epsilon_k}{\sqrt{K}}\right)\sim N(\mu,\sigma^2)$$
\noindent where $\mu$ and $\sigma^2$ are finite for uniformly distributed variables $\epsilon_k$.
This in turn implies that $$\frac{\Sigma_k\epsilon_k}{K}\sim N(\mu,\frac{\sigma^2}{K}).$$
\noindent Hence, the distribution is a $\Delta$ function at $\mu$ for large $K$. The resulting distribution of money
would also be a $\Delta$ function i.e., perfect equality will be achieved. For finite values of $K$ greater than unity,
the distribution of money would resemble a gamma probability density function (see Fig. \ref{div}). $~~~~~~$$\square$

It may be noted that the assumption of the returns having the same mean and variance alongwith independence among 
themselves, 
is not very
realistic. We relax the assumption of independence below but the assumption of the means and the
variances being the same, is maintained throughout the paper as it helps to avoid unnecessary complications. 
Also note that the shifts in the distribution are discrete since
the distribution alters only if the number of pairs of the different commodities traded ($K$)
alters. This is because we have made a strong assumption that the returns viz., $\epsilon_1$, $\epsilon_2$,...,
$\epsilon_K$ etc. are i.i.d. variables as has been mentioned above. 

By allowing the returns to be correlated we can generate continuous shifts in the
distributions instead of discrete jumps for $K=1, 2, 3 ...$ etc. 
For example, consider two specific cases where in the first case, 
there is only one pair of commodities and in the other, two pairs of
commodities to be traded (more precisely, 
we assume $K=1$ and $2$ respectively in Eqn. \ref{diversify}). 
In the first case,
the steady state distribution of money would be a pure exponential 
whereas in the second case, the distribution resembles a gamma pdf 
(see Fig. \ref{div}). By assuming zero correlation we can derive these two limits only. However, if we assume that the
returns may be correlated then by varying the degree of correlation we can get continuous shifts. One noteworthy
feature of this case, is that for a risk-averse agent the risk-minimizing 
choice would be to diversify equally even if there is
any correlation among the random terms.

\section{A Generalized Kinetic Exchange Model}
{
\label{sec:model}

\noindent However, it becomes very difficult to work with the Eqn. \ref{diversify} since 
there are two parameters viz., $K$ (the number of commodities traded) and the correlation among the returns. An
even more troublesome issue is that as $K$ changes, the equilibrium distribution jumps discontinously.
Hence, we simplify Eqn. \ref{diversify} by assuming that there are only two simultaneous trading processes going
on at each instant (that is $K=2$, which essentially implies that there 
are only two random terms; recall the proposition stated earlier).
We modify the model to incorporate the savings propensity and the correlation parameter explicitly
in the trading equations in the following fashion.

\begin{eqnarray}
m_i(t+1) &=& \lambda m_i(t)+\omega_1(1-\lambda) m_i(t)+\left(\alpha\omega_1+(1-\alpha)\omega_2\right) \nonumber \\
&& (1-\lambda) m_j(t) \nonumber\\
m_j(t+1) &=& \lambda m_j(t)+(1-\omega_1)(1-\lambda)m_i(t)+\left(1-\alpha\omega_1-(1-\alpha)\omega_2\right) \nonumber \\
&&(1-\lambda) m_j(t) 
\end{eqnarray}

\noindent which is identical to Eqn. \ref{model}.
Several points are to be noted.

\begin{enumerate}

\item[(a)] If $\lambda=0$ and $\alpha=1$, then we have the very basic framework of ideal gas which gives
rise to a purely exponential distribution (Gibbs distribution: $p(m)\sim e^{-m/T}$ with $T=1$ in this case).
See Ref. \cite{yako-rosser;09, dragu-yako;01}.

\item[(b)] 
\label{barma;00}
If $\lambda=0$ and $\alpha=0$, then we have a model with two uncorrelated stochastic terms. This model
has been studied and solved in Ref. \cite{mbarma;00}. This model gives rise to a probability distribution
characterized by a gamma probability density function of
the form $p(m)\sim 4me^{-2m}$. 

\item[(c)] If $lim ~\lambda\rightarrow 1$, then the distribution would be a delta function.

\item[(d)] If only $\alpha=1$, the above model reduces to the so-called CC model \cite{anirbanc;00} which
gives rise to gamma-function like behavior.

\item[(e)] If only $\alpha=0$, then we have a new model which has savings propensity (CC model) and two
uncorrelated random terms (see point (b) above).
\end{enumerate}

}

\section{Inequality Reversal}
\label{sec:ineq}
{
\noindent Inequality can be measured by a number in indices. However, the most useful one in this case is simply
the coefficient of variation which is basically the standard deviation of the distribution normalized by the
expectation. The economy is modelled in such a way that the expectation is always set equal to unity
(recall that all agents are initially endowed with unit amount of money and the economy under study is
a conserved one i.e., money is neither created nor annihilated in this economy). 
We can calculate the moments recursively
(see Ref. \cite {richmond;05} for more on finding the moments).
We consider the $i$-th agents money evolution equation only.
\begin{equation}
m_i(t+1) = [\lambda+\omega_1(1-\lambda)] m_i(t)+[\alpha\omega_1+(1-\alpha)\omega_2](1-\lambda) m_j(t) 
\label{i-th}
\end{equation}
\noindent 

{\bf Lemma 1}: $\langle m\rangle$=1.\\
{\it Proof}: By taking expectation over both sides of Eqn. \ref{i-th}, 
we get 
$$
\langle m_i(t+1)\rangle = [\lambda+\frac{1}{2}(1-\lambda)] \langle m_i(t)\rangle
+\frac{1}{2}(1-\lambda) \langle m_j(t)\rangle.
$$

Note that $\langle m_j(t)\rangle$= $\sum_{j=1}^{N}m_j(t)/N$=1.
Since the expected money-holding is free of the time index,
the result readily follows.   $~~~~~~~~~~~~~~~~\square$ \\

{\bf Lemma 2}:  
\begin{eqnarray}
\Delta m &=& \frac{2(1-\lambda)
\left[\frac{\alpha(1-\lambda)}{3}+\frac{\lambda}{2}  +
\frac{(1-\lambda)(1-\alpha)}{4}\right]}{1-z}-1
               \nonumber 
\end{eqnarray}
\noindent where $z=(1-\lambda)^2\left(\frac{1}{3}+\frac{\lambda}{(1-\lambda)^2}+\frac{\alpha^2+(1-\alpha)^2}{3}
+\frac{\alpha(1-\alpha)}{2}\right)$.

{\it Proof}: It follows from the definition of variance that
$$\Delta m_i = \langle m_i^2\rangle-(\langle m_i\rangle)^2 \nonumber $$    
where $m_i$ is given by Eqn. \ref{i-th} and $\Delta$ stands for the second central moment (variance). 
By substituting $m_i$ in the {\it l.h.s.} of the expression of variance and noting that $\langle m\rangle=1$, we get
$$
\Delta m_i(t+1)= \langle[ (\lambda+\omega_1(1-\lambda))m_i(t)+(\alpha\omega_1+(1-\alpha)\omega_2)
(1-\lambda)m_j(t) ]^2  \rangle-1.    
$$
Here, we use the fact that in the steady state, the variance of the distribution should be free of the time
and the agent indices. Also, since $\omega_i\sim$ uniform [0,1], $\langle\omega_i\rangle=1/2$ and
$\Delta \omega_i$ = 1/12 (for $i=$1, 2). On simplification, we get
$$
\Delta m=z(\Delta m +1)+2(1-\lambda)\langle\left(\lambda+\omega_1(1-\lambda)\right)\left(\alpha\omega_1
+(1-\alpha)\omega_2\right) \rangle -1       
$$
\noindent where $z=(1-\lambda)^2\left(\frac{1}{3}+\frac{\lambda}{(1-\lambda)^2}+\frac{\alpha^2+(1-\alpha)^2}{3}
+\frac{\alpha(1-\alpha)}{2}\right)$. 
On further simplification, we get
\begin{equation}
\Delta m = \frac{2(1-\lambda)
\left[\frac{\alpha(1-\lambda)}{3}+\frac{\lambda}{2}  +
\frac{(1-\lambda)(1-\alpha)}{4}\right]}{1-z}-1
\label{variance}
\end{equation}

\noindent where $z$ is defined as above.             $~~~~~~~~~~~~~~~~~~~~~~~~~~~~~~~~~~~~~~\square$

\begin{figure}
\begin{center}
\noindent \includegraphics[clip,width= 6cm, angle = 270]
{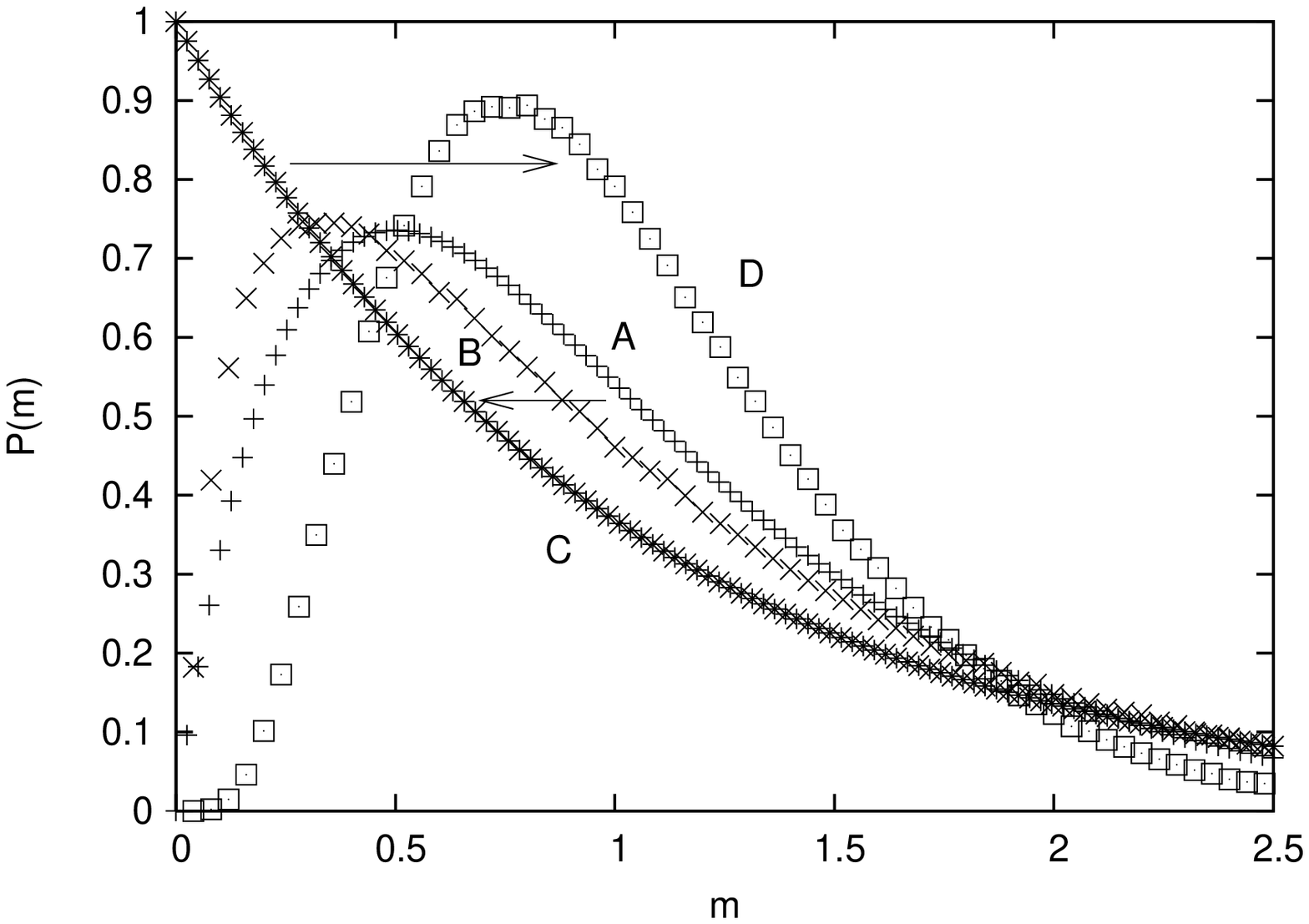}
\caption\protect {Steady state distributions of money for different values
of the parameters. Curve A: $\lambda=0$ and $\alpha=0$. Curve B: $\lambda=0$ and $\alpha=0.7$.
Curve C: $\lambda=0$ and $\alpha=1$. Curve D: $\lambda=0.5$ and $\alpha=1$.
As the correlation goes up the distribution becomes more skewed to the left (from A to B to C; see the arrow). Then as the
savings propensity goes up, it moves in the opposite direction (from C to D; see the arrow). All simulations are done
for $O(10^6)$ time steps with 100 agents and averaged over $O(10^3)$ time steps.}
\label{pdf}
\end{center}
\end{figure}

\begin{figure}
\begin{center}
\noindent \includegraphics[clip,width= 6cm, angle = 270]
{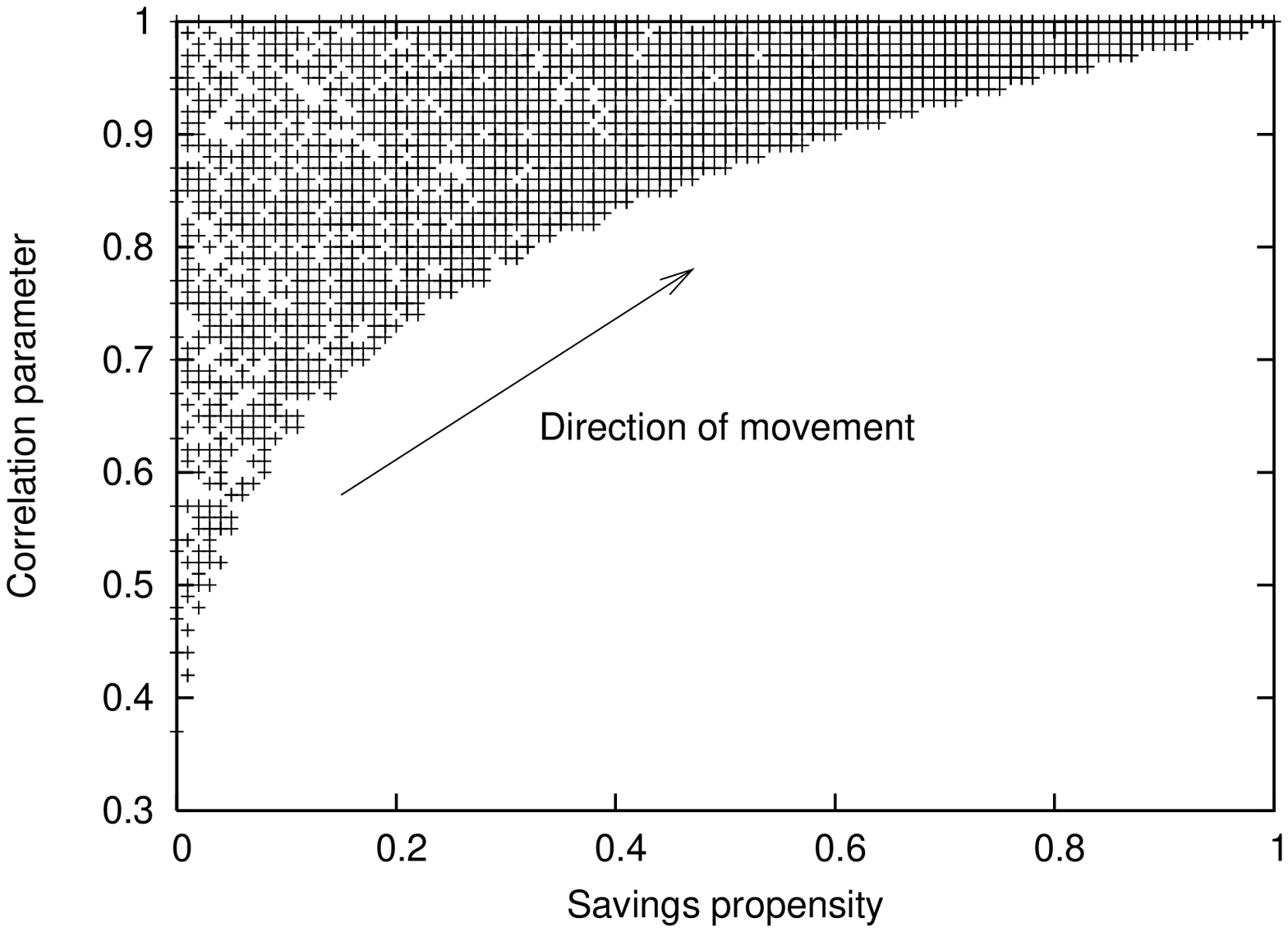}
\caption\protect{The ($\lambda,\alpha$) parameter space. If the economy moves through the
shaded region (the region above the curve $\alpha=\lambda^{1/5}$) then it shows inequality reversal.}
\label{space}
\end{center}
\end{figure}

\begin{figure}
\begin{center}
\noindent \includegraphics[clip,width= 5cm, angle = 270]
{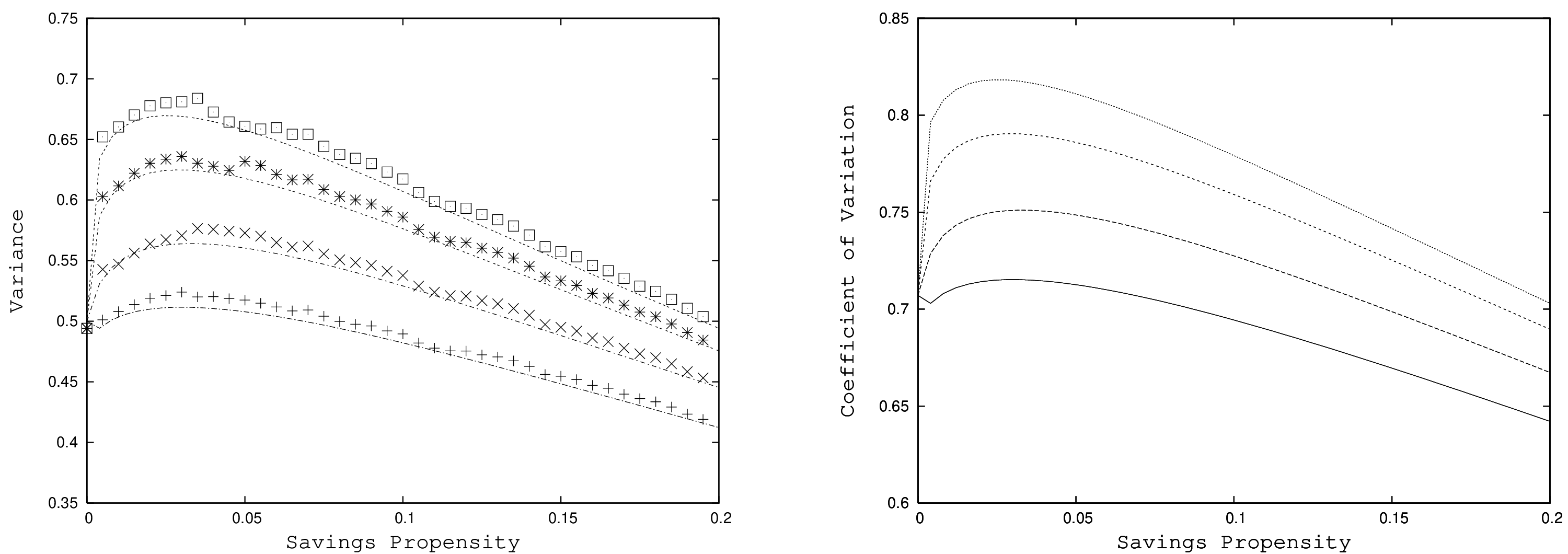}
\caption\protect{{\it Left panel}: The changes in the variance of the distribution ($\Delta m$) is shown with
changes in the savings propensity ($\lambda$) (from below, $\tau$=5, 7, 10 and 13). The Monte Carlo simulation results agree
with the theoretical curves (dotted lines) obtained from the Eqn. \ref{variance} and \ref{tau}. {\it Right panel}:
The Kuznets Curve in terms of the Coefficient of Variation i.e., $\sqrt{\Delta m}$
(from below, $\tau$=5, 7, 10 and 13): inequality increases and then falls.}
\label{curve}
\end{center}
\end{figure}

\begin{figure}
\begin{center}
\noindent \includegraphics[clip,width= 6cm, angle = 270]
{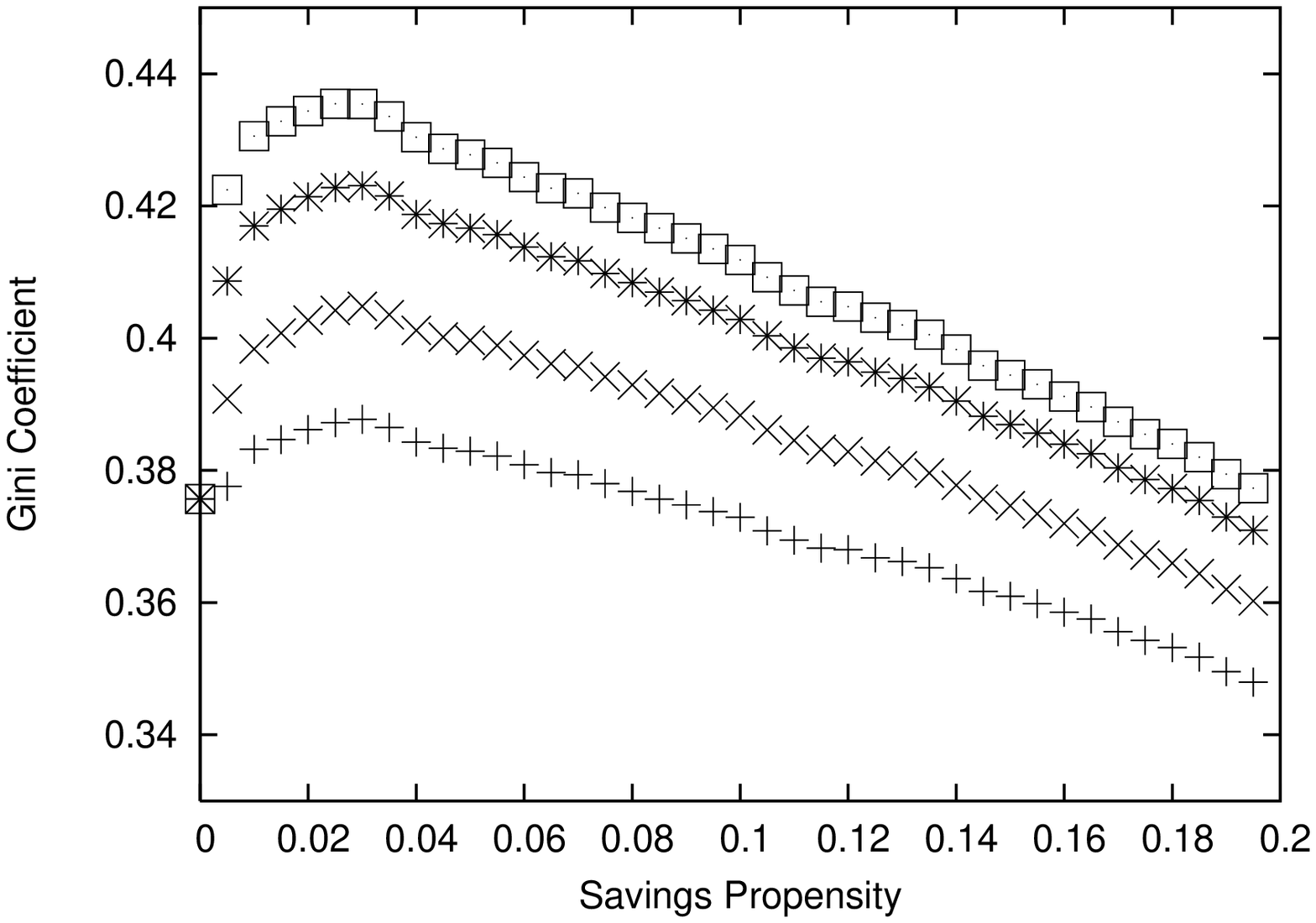}
\caption\protect{The Kuznets Curve in terms of the Gini concentration ratio (from below, $\tau$=5, 7, 10 and 13):
inequality increases and then falls. All Monte Carlo simulations are done for $O(10^5)$ time steps with 100 agents and
averaged over $O(10^2)$ time steps.}
\label{gini}
\end{center}
\end{figure}

Clearly the variance is a function of $\lambda$ and $\alpha$ only. 
Now, we make use of two observations. First, for a sustainable growth the savings propensity 
has to increase (see Ref. \cite{sala-i-martin} for more on this topic). 
The second observation is that the modern markets are characterized by
correlated returns with fluctuations \cite{sornette} in the most efficient state \cite{bak}. 
The implications are that both $\lambda$ and $\alpha$ increases over time unidirectionally. 
By plugging different values of $\lambda$ and $\alpha$ in the expression of variance, one can find how
inequality changes over time with increases in the parameters.
Note that since the parameters are ranging between $0$ and $1$, the parameter space is a square with unit length 
(Fig. \ref{space} shows the relevant region). 
Now, we assume that the path followed by the economy starts from the origin ($\lambda=0$, $\alpha=0$) and
ends at ($\lambda=1$, $\alpha=1$). The simplest functional relationship between $\lambda$ and $\alpha$
satisfying the above assumption is of the form 
\begin{equation}
\alpha=\lambda^{\frac{1}{\tau}} 
\label{tau}
\end{equation}
\noindent where $\tau$ is a positive number. It is numerically seen that for $\tau \geq 5$, the economy
shows a very prominent inequality reversal (see Fig. \ref{curve}).
It should be noted, however, that if the economy follows some other paths in the parameter space, then it
may show other types of behavior as well.

For the sake of completeness, we also provide Monte Carlo simulation results of the
Kuznets Curves where the inequality is measured in terms
of the Gini concentration ratio. The definition of the measure $G$ is the following \cite{kleiber}.

\begin{equation}
G\equiv \frac{\sum_{i=1}^{N}\sum_{j=1}^{N} \mid m_i-m_j\mid}{2 \mu N(N-1)}   
\end{equation}

where $N$ is the number of agents (which is set to 100), $\mu$ is the money per agent (which is set to unity) 
and $m_i$ is the money holding of the $i-$th agent.
Fig. \ref{gini} shows the rise and the subsequent fall in the Gini concentration ratio.

In its original formulation, the Kuznets Curve is a plot of income distribution with the changes in income
per capita. However, the trading process that we have considered is a conservative one, implying that the
average income in this model remains fixed over time. Instead of average income, we consider the changes
in the savings propensity and the correlation among the markets and we trace the corresponding changes in the
inequality in money-holding. In such cases, it is clearly seen that the economy shows Kuznets-type behavior.

\section{Summary}
{
\noindent The presence of inequality is a persistent phenomena in any economy. The static nature of inequality
has been under investigation for more than a decade \cite{acbkc;07, yako-rosser;09}.
Here in this paper, we examine the dynamic aspects of inequality. It is seen that as an economy grows
over time, its inequality first increases and then falls with further growth. This particular dynamic feature
was first pointed out by Kuznets \cite{kuznets;55,kuznets;65}. 
This observation has attracted considerable interests in Economics. In this paper, we have tried to explain the origin
of such a phenomena in an appropriately modified kinetic exchange model (see Ref. \cite{acbkc;07,
chakrabartis;10} for recent reviews on kinetic exchange models).

First, following Ref. \cite{chakrabartis;09} we propose a basic binary trading 
process and we derive the corresponding money transfer equations
as the competitive solution the trading action.
Then we propose another
model to show the effects of correlated returns on the equilibrium distributions of money.
Here, we consider the case where each agent produces a
number of commodities
and hence, would diversify their money holding in trading of different commodities to minimimize
the risk (Sec. \ref{sec:corrmkt}).
Then we formulate a very general kinetic market model incorporating the effects of savings propensity
and correlated returns (Sec. \ref{model}) explicitly. 
In different limits, this model gives rise to a number of kinetic exchange 
models viz., the Dragulescu-Yakovenko model \cite{dragu-yako;01},
Chakraborti-Chakrabarti model \cite{anirbanc;00} and yet another model proposed in Ref. \cite{mbarma;00}.
It is observed that with the growth of an economy, the savings propensity of its
people increases \cite{sala-i-martin} and the efficient markets are characterized by correlated returns from trade 
\cite{sornette, bak}. Using these observations,
we show that if the economy
moves through a certain region in the parameter space then it shows inequality reversal as was found by 
Kuznets \cite{kuznets;55, kuznets;65} (Sec. \ref{sec:ineq}). The Monte Carlo
simulation results are also explained by finding out
the expression of the inequality index as well (Eqn. \ref{variance}).

}


\end{document}